# Towards "Intelligent Compression" in Streams: A Biased Reservoir Sampling based Bloom Filter Approach

Sourav Dutta
IBM Research, India
New Delhi
sodutta3@in.ibm.com

Souvik Bhattacherjee
IBM Research, India
New Delhi
souvikbh@in.ibm.com

Ankur Narang
IBM Research, India
New Delhi
annarang@in.ibm.com

## ABSTRACT

With the explosion of information stored world-wide, data intensive computing has become a central area of research. Efficient management and processing of this massively exponential amount of data from diverse sources, such as telecommunication call data records, telescope imagery, online transaction records, web pages, stock markets, medical records (monitoring critical health conditions of patients), climate warning systems, etc., has become a necessity. Removing redundancy from such huge (multi-billion records) datasets resulting in resource and compute efficiency for downstream processing constitutes an important area of study. "Intelligent compression" or deduplication in streaming scenarios, for precise identification and elimination of duplicates from the unbounded data stream is a greater challenge given the real-time nature of data arrival. Stable Bloom Filters (SBF) address this problem to a certain extent. However, SBF suffers from a high false negative rate (FNR) and slow convergence rate, thereby rendering it inefficient for applications with low FNR tolerance.

In this paper, we present a novel *Reservoir Sampling based Bloom Filter*, ($RSBF$) data structure, based on the combined concepts of *reservoir sampling* and *Bloom filters* for approximate detection of duplicates in data streams. Using detailed theoretical analysis we prove analytical bounds on its false positive rate ($FPR$), false negative rate ($FNR$) and convergence rates with low memory requirements. We show that $RSBF$ offers the currently lowest FN and convergence rates, and are better than those of SBF while using the same memory. Using empirical analysis on real-world datasets (3 million records) and synthetic datasets with around 1 billion records, we demonstrate upto 2× improvement in FNR with better convergence rates as compared to SBF, while exhibiting comparable FPR. To the best of our knowledge, this is the first attempt to integrate reservoir sampling method with Bloom filters for deduplication in streaming scenarios.

## 1. INTRODUCTION

Data intensive computing has evolved into a central theme in the research community and the industry. There has been a tremendous spurt in the amount of data being generated across diverse application domains such as IR, telecommunication (call data records), online transaction records, web pages, medical records, virus databases and climate warning systems to name a few. Processing such enormous data is computationally prohibitive, and is further compounded by the presence of duplicates and redundant data, wasting precious compute time. Removing redundancy in the data helps in improving resource utilization and compute efficiency especially in the context of stream data, which generally requires real-time processing at 1 GB/s or higher. In this work, we consider the problem of real-time elimination of redundant records present in large streaming datasets. A record may be considered redundant, if it had arrived previously in the stream. Formally, this is referred to as the *data deduplication* or *intelligent compression* problem. Data redundancy removal (DRR) and deduplication are used interchangeably in this paper.

Consider, for example, a large nation wide telecommunication network, where each call generates call data records (*CDRs*). Each CDR contains details about a particular call such as the calling number, the called number and so forth. Due to errors in CDR generation, multiple copies of a CDR may get generated. Before storing these CDRs in a central data center, one needs to perform deduplication over around 5 billion CDRs at real-time performance. Solutions involving database accesses as in traditional systems are prohibitively slow. Since algorithms involving typical Bloom filters such as [11] are extremely resource intensive with huge memory requirements (20GB or higher for 6B CDRs at FPR = $1e − 5$), applications have to resort to disk based Bloom filer data structures at the expense of reduced performance. Hence, there is a strong need for deduplication algorithms that work in-memory or with reasonable memory, have real-time performance and also have low FPR, FNR and better convergence rates. This poses a very challenging problem.

Search engines regularly crawl the Web to update their corpus of webpages. Given the list of URLs extracted from the content of a crawled page, a search engine must probe its archive to determine if the URL is already present in its collection, and if re-crawling of the URL can be avoided [?]. This involves duplicate detection, which in practice may be imprecise but is indispensable given the number of webpages present in the Internet. The consequence of an imprecise duplicate detection is that some already-crawled pages will be crawled again (caused by FNR), or some new URLs which should be crawled are missed (caused by FPR). Here, a high FNR might lead to severe performance degradation, while a relatively high FPR results in new pages being ignored leading to a stale corpus of webpages, both of which need to be balanced since a search engine can archive only a small portion of the entire web [6].

[17] proposes another application for approximate duplicate detection in a streaming environment. In a Web advertising scenario,

advertisers pay web site publishers for clicks on their advertisements. For the sake of profit, it is possible that a publisher fakes some clicks (using scripts). Hence a third party, the advertising commissioner, has to detect those false clicks by monitoring duplicate user IDs and IPs. Here, low FNR is necessary to ensure minimal fraud. Ensuring low FNR while simultaneously having low FPR, along with memory efficiency presents a difficult scenario.

Straightforward approaches for data redundancy removal (DRR) involve pair-wise string comparisons, leading to quadratic complexity. This prohibits real-time redundancy removal over enormous (1 to 10 billion) number of records. In order to address this computational challenge, Bloom filters [4] are typically used. Bloom filters are space-efficient probabilistic data structures that provide fast set membership queries, but with a small false positive rate (*FPR*). Parallel Bloom filter based algorithms have also been explored in [11].

Typical Bloom filter approaches involve $k$ comparisons for every record, where $k$ is the number of hash functions computed per record to check the bits of the Bloom filter array. This leads to poor performance, as for an in-memory DRR over billions of records, the memory required by such Bloom filter array is very high (order of tens of Gigabytes depending on the false positive rate). One approach is to store the Bloom filter array on the disk and bring parts of it into memory for reading and updates. But, this would lead to a huge fall in the overall DRR throughput (due to disk access overheads). Further, there is a trade-off between the cache performance and memory efficiency [19] in such Bloom filter design.

In order to address these challenges, we present the design of a novel Bloom filter based on biased Reservoir Sampling [22, **?**], referred to as *RSBF (Reservoir Sampling based Bloom Filter)*. Using threshold based non-temporal bias function we obtain upto $2\times$ improvement in FNR and much better convergence rates as compared to [6] while maintaining nearly the same FPR. The choice of such bias functions may be of independent research interest in this direction.

This paper makes the following contributions:

(1) We present the design of a novel Bloom filter based on biased Reservoir Sampling, *RSBF*. Using threshold based non-temporal bias function, we obtain improved FNR and convergence rates as compared to [6] while maintaining similar FPR.

(2) Using detailed theoretical analysis, we provide upper bounds on FPR and FNR. Further, we exhibit the faster convergence of our algorithm, compared to *SBF*, with expected bounds on the number of 1s in the Bloom filters.

(3) We demonstrate real-time in-memory DRR using both real and synthetic datasets of the order of 1B records. We observe upto $2\times$ better FNR and much better convergence rates compared to the prior results.

## 2. PRELIMINARIES & BACKGROUND

A Bloom filter is a space-efficient probabilistic data structure that is widely used for testing membership queries on a set [3]. The efficiency is achieved at the expense of a small false positive rate, where the Bloom filter may falsely report the presence of an element in the set. However, it does not report false negatives, i.e. falsely reporting the absence of an element in the set. Representing a set of $n$ elements by a Bloom filter requires an array of $m$ bits ($m << n$), initially all set to 0. To insert an element $e_i$ into the Bloom filter, $k$ bits (locations) in the Bloom filter array are set. These $k$ locations are evaluated from $k$ independent hash functions $h_1(e_i), \ldots, h_k(e_i)$. If all the locations are already set to 1, then either the element $e_i$ is already a member of the set or is a false positive. The probability of the false positive rate [4] for a standard Bloom filter is given by:

$$FPR \approx \left(1 - e^{-kn/m}\right)^k \qquad (2.1)$$

Given $n$ and $m$ the optimal number of hash functions $k = \ln 2.(m/n)$. For detailed analysis of these derivations, please refer [4].

To support a situation where the contents of a set changes over time, with elements being continually inserted and deleted, Fan et al. [9] introduced counting Bloom filters. This approach allows elements to be updated in the Bloom filter by using a small counter instead of a single bit at every position. Insertion now requires the corresponding counters to be incremented. On the other hand, deletion requires the corresponding counters to be decremented. The process of deletion introduces a *false negative*, (*FN*) wherein an element is wrongly reported as unique.

In reservoir sampling [22], one continuously maintains a reservoir of size $n$ from the data stream. The first $n$ points in the data stream are added to the reservoir for initialization. Subsequently, after $t$ elements of the data stream have been processed, the $(t+1)^{\text{th}}$ element is added to the reservoir with probability $n/(t + 1)$, also known as the *insertion probability*. This element replaces a randomly chosen element from the current reservoir. We note that the probability value $n/(t + 1)$ reduces with stream progression. Reservoir sampling thus satisfies the following property:

**Property** After $t$ points in the data stream have been processed, the probability of any point in the stream belonging to the sample of size $n$ is equal to $n/t$.

One interesting characteristic of this maintenance algorithm is that it is extremely efficient to implement in practice. When new points in the stream arrive, we only need to decide whether or not to insert into the current sample array which represents the reservoir. The sample array can then be overwritten at a random position. The bias function [**?**] associated with the $r^{\text{th}}$ data point at the time of arrival of the $t^{th}$ point ($r \le t$) is given by $f(r, t)$ and is related to the probability $p(r, t)$ of the $r^{\text{th}}$ point belonging to the reservoir at the time of arrival of the $t^{\text{th}}$ point. Specifically, $p(r, t)$ is proportional to $f(r, t)$. The function $f(r, t)$ is monotonically decreasing with $t$ (for fixed $r$) and monotonically increasing with $r$ (for fixed $t$). Therefore, the use of a bias function ensures that recent points have higher probability of being represented in the sample reservoir. Hence, we define the concept of a bias-sensitive sample $S(t)$, which in turn is defined by the bias function $f(r, t)$ as,

**Definition** Let $f(r, t)$ be the bias function for the $r^{\text{th}}$ point at the arrival of the $t^{th}$ point. A biased sample $S(t)$ at the time of arrival of the $t^{th}$ point in the stream is defined as a sample such that the relative probability $p(r, t)$ of the $r^{\text{th}}$ point belonging to the sample $S(t)$ (of size $n$) is proportional to $f(r, t)$.

## 3. RELATED WORK

Duplicate detection poses a classical problem within the domain of data storage and databases giving rise to numerous buffering solutions. With the advent of online arrival of data and transactions, detection of duplicates in such streaming scenarios using similar buffering and caching mechanisms [**?**] constitutes a naïve solution given the inability to store the entire information arriving in an infinite stream. Hence fuzzy duplicate detection methods [**?**, **?**] present an alternative method for tackling the problem.

Data stream management has emerged as a fundamental research domain involving approximate frequency moments [**?**], element classification [**?**], correlated aggregate queries [**?**] to name a few. *Bit Shaving*, the problem of fraudulent advertisers not paying commission for a certain fraction of its traffic has been studied in [**?**].

Approximate duplicate detection has been an area of concern both in the domain of database management and Web applications. Algorithms for redundancy removal for search engines were studied in [?, ?, ?]. File-level hashing was used in storage systems to detect duplicates [?, ?, ?]. However this techniques provides a low compression ratio. For fixed-sized data blocks, [?] proposed secure hashes. Bloom filter was first used by the TAPER system [?]. Further caching techniques have also been applied on Bloom filters as in [?].

In this paper we put forth a novel approximate deduplication algorithm in streaming environments using Bloom filters [3]. The literature contains several proposed Bloom filter variants to suit various application needs for deduplication. These include, counting Bloom filters [9], compressed Bloom filters [18], space-code Bloom filters [15], and spectral Bloom filters [21] among many. Counting Bloom filters replace an array of bits with counters in order to count the number of items hashed to a particular location. The others use subtle variations to efficiently meet the nature of demand of the applications.

The window model of Bloom filters [17] also contains several flavors such as landmark window, jumping window, along with the recently proposed sliding window [?], all of which operate on a definite amount of history of objects observed in the stream to draw conclusions for future processing of the stream elements.

Another exciting Bloom filter structure proposed recently, $SBF$ [6] provides a stable guarantee regarding the nature of performance of the structure given a very large stream. This constant performance is of huge importance in real-time applications involving de-duplication. It continuously evicts stale information from the Bloom filter to make room for more recent elements. It also provides a tight upper bound of false positive rates, however theoretically it attains stability at infinite stream length. In this paper, we used Biased Reservoir sampling based Bloom filter and prove upper bounds on both FPR, FNR and fast convergence to stability. Using empirical analysis, we demonstrate around $2\times$ better FNR compared to [6] and also better convergence rates.

Interestingly, Bloom filters have also been applied to network-related applications, albeit for solving different problems, such as finding heavy flows for stochastic fair blue queue management [10], providing a useful tool to assist network routing, such as packet classification [2], per-flow state management and the longest prefix matching [7]. [13] proposes a new Bloom filter structure that supports representation of items with multiple attributes and exhibits a low false positive rate. It is composed of multiple Bloom filters and a hash table to represent items accurately and efficiently. [12] extends *Bloomjoin*, the state-of-the-art algorithm for distributed joins, to minimize the network usage for the query execution based on database statistics. [16] discusses how Bloom filters can be used to speed up name-to-location resolution process in large scale distributed systems.

A related problem of finding the number of distinct elements present in a data stream was explored in [?]. There exists several other methodologies in the data stream domain to approximate the frequency and norms of the input elements. Our approach as presented in this paper provides a conjugation of Bloom filters and reservoir sampling technique to efficiently approximate duplicate detection in such unbounded streams.

The problem of synopsis maintenance [?] [?] has been studied in great detail due of its extensive application for query estimation [?] in data streams. Many synopsis methods such as sampling, wavelets, histograms and sketches are designed for use with specific applications such as approximate query answering. A comprehensive survey of stream synopsis construction algorithms may be found in [1]. An important class of stream synopsis construction methods is *reservoir sampling* [22]. The method of sampling has great appeal because it generates a sample of the original multi-dimensional data representation. Hence, it can be used with arbitrary data mining applications with minor changes to the underlying methodologies and algorithms.

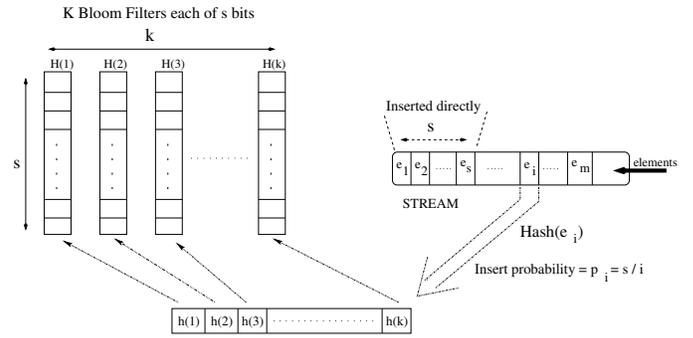

**Figure 1: The structure of RSBF**

[?] proposes a new approach on memory-less temporal bias function based reservoir sampling for continually evolving data streams. It demonstrates that such bias functions lead to efficient implementation: $O(1)$ processing time per stream element. While biased reservoir sampling is a difficult problem (with the one pass constraint), [?] shows that it is possible to design very efficient replacement algorithms for such important class of "memory-less" bias functions. In addition, incorporation of bias results in upper bounds on reservoir sizes in many cases limits the maximum space requirements to nearly constant even for an infinitely long data stream. This enables its application in a variety of space-constrained scenario.

In this paper, we present a non-temporal threshold based bias function for reservoir sampling for the deduplication problem using Bloom filters, resulting in low FNR. Further, we establish theoretically and show empirical results to support the efficiency and fast convergence rates of our algorithm.

## 4. RESERVOIR SAMPLING BASED BLOOM FILTER APPROACH (RSBF)

The design of $RSBF$ (Figure 1) is motivated by the reservoir sampling technique [22] and is targeted for detecting duplicates in large data streams. We consider $k$ Bloom filters each of size $s$ bits. Initially all the bits are set to zero. Each element of the stream is mapped to one of the $s$ bits in each of the $k$ different Bloom filters. Each of these $k$ bits is generated by a uniform random hash function. These $k$ locations are also probed to determine whether the element is *distinct* or *duplicate*, similar to the procedure followed in regular Bloom filters. If all the $k$ bits are set to 1, then the element is said to be *duplicate*, otherwise *distinct*.

The initial $s$ elements of the data stream are directly inserted into the $RSBF$, as in reservoir sampling method, by setting the corresponding $k$ bits in the Bloom filters. Each element $e_i$, $i > s$ of the stream is then inserted with a probability $p_i = s/i$ (*insert probability*), where $i$ is the current length of the stream and $s$ is the size of a Bloom filter (the reservoir). For inserting the element $e_i$, the $k$ hash bits are generated by the hash functions and the corresponding Bloom filter bits are set to 1.

In order to accommodate future elements in the infinite stream within a limited memory space and simultaneously prevent high false positive rates, whenever an element is inserted, we reset $k$ bits to 0. These $k$ bit locations, one from each of the Bloom filters are chosen uniformly at random. However, this deletion operation leads to the occurrence of false negatives in the structure. We observe that the use of reservoir sampling technique in the Bloom filters increases the probability that an element will not be inserted (due to a possible duplicate) into the Bloom filters with decrease in the insert probability as the stream progresses. As a result, the false negative rate of the structure increases, since even new elements found later in the stream may be repeatedly rejected by the reservoir algorithm, thereby degrading the performance of $RSBF$.

To address this problem, we propose a novel extension which can be considered as a weak form of biased reservoir sampling performed on the stream. After the reservoir sampled insertion probability falls below a specific threshold, say $p^*$, any element in the stream reported as unique by probing its corresponding bits in the $k$ Bloom filters, is inserted. This helps to keep the false negative rate in check as the next time the same element arrives, it will correctly be reported as a duplicate. This procedure also enables the Bloom filters to evolve with changes in the data skew of the stream, and help $RSBF$ to dynamically adapt itself to a changing stream. The reservoir sampling still operates on the other elements of the stream.

As the stream length increases, the probability that an element is duplicate increases (for finite universe of the stream elements). If most of the elements in the Bloom filters become 1, the false positive rate increases. The reservoir sampling method, helps to prevent such a scenario by rejecting elements. While this leads to increase in the false negative rate, the use of the threshold $p^*$, helps to control the rate in $FNR$. Thus this novel combination of reservoir sampling and the threshold, complements each other; keeping both the $FPR$ and the $FNR$ at acceptably low limits. The pseudo-code for the working of $RSBF$ is given in Algorithm 1.

We emphasize that the insertion procedure of $RSBF$ selects $k$ bits (one bit for each of the Bloom filters) to be set to 1 and another $k$ bits to be reset to 0. This approach leads to a near constant number of 1s and 0s in $RSBF$, *stability* as discussed later in the paper. Hence, $RSBF$ exhibits significantly lower FNR and faster convergence rate to stability with comparable FPR as that of $SBF$, making it a more attractive structure for modern day applications. In the remaining paper we describe in details and validate with theoretical bounds and empirical results the efficient performance of $RSBF$.

## 5. THEORETICAL FRAMEWORK

In this section, we present theoretical bounds and analysis for FPR, FNR and the fraction of ones (convergence rates) of our RSBF data structure. Later we present existence results justifying the validity of our approach. Table 1 describes the symbols used throughout the paper.

### 5.1 False Positive Rate

Here we compute the false positive rate, $(FPR)$ of our proposed algorithm. A false positive, $FP$ occurs when a distinct element of the stream is reported as a duplicate.

Consider the FPR at $e_{m+1}$, the $(m + 1)^{th}$ element of the stream. We assume that the elements of the stream are uniformly drawn at random from a universe $\Gamma$, with $|\Gamma| = U$.

Let $P_{unique}$ be the probability that $e_{m+1}$ has not occurred in the first $m$ elements of the stream.

$$P_{unique} = \left(\frac{U-1}{U}\right)^m \qquad (5.1)$$

---

**Algorithm 1:** $RSBF(S)$

**Require:** Threshold FPR ($FPR_t$), Memory in bits ($M$), and Stream ($S$)

**Ensure:** Detecting *duplicate* and *distinct* elements in $S$

Compute the value of $k$ from $FPR_t$.
Construct $k$ Bloom filters each having $M/k$ bits of memory.
$iter \leftarrow 1$
$flag \leftarrow 0$
**for** each element $e$ of $S$ **do**
  Hash $e$ into $k$ bit positions, $H = h_1, \cdots, h_k$.
  **for** each $h_i$ in $H$ **do**
    **if** at position $h_i$ in the $i^{th}$ bloom filter is not set **then**
      $Result \leftarrow DISTINCT$
      $flag \leftarrow 1$
      break
    **end if**
    **if** flag = 0 **then**
      $Result \leftarrow DUPLICATE$
    **end if**
    Compute probability of insertion of $e$, $P_e$
    **if** ($P_e \leq (s/iter)$) OR ($iter \leq s$) **then**
      **for** all positions $h_i$ in $H$ **do**
        Set the bit at $h_i$ of the $i^{th}$ bloom filter.
      **end for**
    **else**
      **if** ($P_e > p^*$) AND ($Result = DISTINCT$) **then**
        **for** all positions $h_i$ in $H$ **do**
          **if** $h_i = 0$ **then**
            Find a bit in $i^{th}$ bloom filter which is set to 1, and reset to 0.
            Set the bit at $h_i$ position to 1
          **end if**
        **end for**
      **else**
        No operation.
      **end if**
    **end if**
    $iter \leftarrow iter + 1$
  **end for**
**end for**

---

Let $e_{m+1}$ hash to $H = \{h_1, h_2, \ldots, h_k\}$ positions, where $h_i \in [1, s]$ for the $i^{th}$ Bloom filter. $e_{m+1}$ will be reported as a duplicate when all the bit positions in $H$ are set to 1 after the first $m$ stream elements. Since all the Bloom filters are identical and are independently processed, we argue with one of them, and then extended for the others.

Assume element $e_l$ hashes to position $h_1$ in the first Bloom filter. Initially, all the bits of the Bloom filters are set to 0. Let the latest transition of $h_1$, from 0 to 1, occur at the $l^{th}$ iteration, and thereafter $h_1$ is never reset, i.e. set to 0. We observe, in $RSBF$ a bit will not be reset to 0 in any iteration, if the stream element for the iteration is not selected for insertion or a different bit of the Bloom filter is chosen for deletion, if the element is to be inserted. We represent the probability of such a transition of $h_1$ by $P_{trans}$. Therefore,

$$P_{trans} = P(e_l \text{ is inserted}).P(e_l \text{ selects } h_1).P(h_1 \text{ is not reset})$$

$$= p_l.\frac{1}{s}.\prod_{i=l+1}^{m}\left[(1-p_i)+p_i.\frac{s-1}{s}\right] = \frac{s}{l}.\frac{1}{s}.\prod_{i=l+1}^{m}\left(1-\frac{1}{i}\right) = \frac{1}{m}$$

| Symbols | Meanings |
|---------|----------|
| M | Available memory (in bits) |
| k | Number of bloom filters |
| s | Size of each bloom filter (in bits) |
| $p_i$ | Prob. of insertion by Reservoir Sampling |
| $p^*$ | Insertion threshold prob. for distinct elements |
| $h_i$ | Hash position within the $i^{th}$ bloom filter |
| $S$ | Stream of input elements |

**Table 1: Symbol List**

This transition may happen during any of the iterations from $(s+1)$ to $m$. Hence, $l \in [s+1, m]$ giving,

$$P_{range} = \sum_{l=s+1}^{m} P_{trans} = \sum_{l=s+1}^{m} \frac{1}{m} = \frac{m-s}{m} \quad (5.2)$$

Since the different Bloom filters are independent, the analysis for other bit positions in $H$ hold similarly as given by Eq. (5.2). The final decision (*distinct or duplicate*) regarding $e_{m+1}$ is taken after probing all the bit positions of $H$, and hence,

$$P_H = \left(1 - \frac{s}{m}\right)^k \approx \left(1 - \frac{k.s}{m}\right) \quad (5.3)$$

It can be observed that the transition of the bit may also be possible during the first $s$ elements of the stream. Since the first $s$ elements of the stream are always inserted, all the bit positions in $H$ should be set at least once during this period for an element to be reported as duplicate. Therefore, the probability that the bit in a Bloom filter is set in the initial $s$ iterations is given by,

$$P_{s\_set} = 1 - \left(\frac{s-1}{s}\right)^s \approx \left(1 - \frac{1}{e}\right) \quad (5.4)$$

This bit must not be reset during the $(s+1)^{th}$ to $m^{th}$ iterations, which is given by,

$$P_{reset'} = \prod_{i=s+1}^{m} \left[p_i.\left(1 - \frac{1}{s}\right) + (1 - p_i)\right] = \frac{s}{m} \quad (5.5)$$

Using Eqs. (5.4) and (5.5), the probability of all the bits for $e_{m+1}$ at positions in $H$ being set is,

$$P_{H_s} = \left(\left[1 - \frac{1}{e}\right].\frac{s}{m}\right)^k \quad (5.6)$$

Either of the above two events will contribute to the $FPR$, hence the probability of $e_{m+1}$ being reported as an $FP$ can be obtained by using Eqs. (5.1), (5.3), and (5.6), which is given by,

$$P_{FPR} = \left(\frac{U-1}{U}\right)^m . \left[1 - \frac{k.s}{m} + \left(\left[1 - \frac{1}{e}\right].\frac{s}{m}\right)^k\right] \quad (5.7)$$

Analyzing Eq. (5.7), we observe that as the stream length $m$ tends to infinity, the right multiplicative factor tends to 1. However, as $U - 1 < U$ the left multiplicative term tends to 0. Hence as the stream length increases, the observed $FPR$ decreases and nearly becomes constant. This leads to a stable performance of $RSBF$ similar to that of $SBF$. However, $RSBF$ achieves this convergence much faster as opposed to $SBF$ as discussed in Section 5.3. In Section 6, we exhibit extensive experimental results that validates this claim.

## 5.2 False Negative Rate

A false negative (FN) error occurs in a stream when a duplicate element is recognized as distinct. In this section, we focus on determining the probability of occurrence of an FN. As per the working of $RSBF$ (Algorithm 1), an element $e$ will be an FN if it has occurred in the stream earlier and one of the following two cases hold:

1. At least one of the $k$ bits of the hash positions of $e$ (set during the previous occurrence of $e$) has been reset during the insertion of another stream element into the reservoir.

2. When $e$ occurred earlier in the stream it was not inserted due to low insertion probability of the stream then (by Reservoir Sampling). However, according to the threshold $p^*$ in Algorithm 1, we insert every distinct element in the stream if the current insertion probability, $p_i \leq p^*$. Therefore, if previous appearances of $e$ had occurred before $p_i$ was less than $p^*$ and were not inserted, then it is likely to be detected as an FN when $e$ repeats for the first time after the insertion probability of the reservoir falls below $p^*$.

We now consider the probability of occurrence of an FN for an element $e_{m+1}$, at the $(m + 1)^{th}$ iteration. Let the previous occurrence of element $e_{m+1}$ be at position $x$, where it was inserted into the reservoir. Therefore, $\Pr(e_{m+1}$ occurs at $x$ and is inserted) $= P_x = p_x/U$. Now, for all iterations from $(x + 1)$ to $m$, either $e_{m+1}$ has not occurred in the stream or was not inserted. Thus, $\Pr(e_{m+1}$ has not occurred OR $e_{m+1}$ has not been inserted after $x$) is given by,

$$P_{x'} = \prod_{i=x+1}^{m} \left[\frac{U-1}{U} + \frac{1-p_i}{U}\right] \leq \left[1 - \frac{s}{U.m}\right]^{m-x} \quad [\because p_i = s/i]$$

$$\leq e^{\frac{-s(m-x)}{U.m}} \quad [\because \frac{s}{U.m} \text{ is small}] \quad (5.8)$$

Now, $\Pr(e_{m+1}$ was last inserted at position $x)$ is given by,

$$P_l = P_x \cdot P_{x'} \leq \frac{s}{U.x} \cdot e^{\frac{-s(m-x)}{U.m}} \quad (5.9)$$

Since $e_{m+1}$ was last inserted at position $x$, the $k$ bits corresponding to $e_{m+1}$ were all set to 1. Therefore, $e_{m+1}$ will be a FN if at least one of those $k$ bits is reset to 0. Due to the deletion operation in case of insertion of an element into the reservoir, some of those $k$ bits can be reset again. Let $y$ be the last iteration where there is a transition from 0 to 1 for any of the $k$ n=bits corresponding to $e_{m+1}$, after which it is not reset again till the $m^{th}$ iteration, and hence $x \leq y \leq m$. Therefore, $\Pr(a$ bit is set at $y) = P_y = p_y/s = 1/y$. Also, $\Pr(that$ bit is not reset after $y)$ is given as

$$P_{y'} = \prod_{i=y+1}^{m} \left[p_i \left(1 - \frac{1}{s}\right) + (1 - p_i)\right] = \frac{y}{m} \quad (5.10)$$

Hence, $\Pr(the$ last transition of the bit from 0 to 1 in a buffer at $y)$ can be expressed as a product of $P_y$ and $P_{y'}$ which is equal to $1/m$.

As $y$ can vary from $x$ to $m$, therefore the $\Pr($ the bit remains set at $m) = \sum_{y=x}^{m} \frac{1}{m} = \frac{m-x+1}{m}$. So, the $\Pr(at$ least one of those $k$ bits is reset at $m) = P_r = 1 - \Pr(the$ bit remains set at $m)^k = 1 - \left(\frac{m-x+1}{m}\right)^k$. Now, $\Pr(e_{m+1}$ is last inserted at $x$ AND at least one of those $k$ bits is reset at $m)$ is given by

$$P_{lr} = P_l \cdot P_r \leq \frac{s}{U.x} \cdot e^{\frac{-s(m-x)}{U.m}} \cdot \left[1 - \left(\frac{m-x+1}{m}\right)^k\right] \quad (5.11)$$

Let

$$\gamma = s \cdot e^{\frac{-s(m-x)}{U.m}} = \frac{s}{1 + \frac{s(m-x)}{U.m} + \frac{\left[\frac{s(m-x)}{U.m}\right]^2}{2!} + \cdots}$$

$$= \frac{1}{\frac{1}{s} + \frac{m-x}{U.m} + \frac{s(m-x)^2}{2!(U.m)^2} + \cdots} \leq 1 \qquad (5.12)$$

Therefore, substituting Eq. (5.12) in Eq. (5.11) we have,

$$P_{lr} \leq \frac{1}{U.x} \cdot \left[1 - \left(\frac{m-x+1}{m}\right)^k\right] \qquad (5.13)$$

However, the value of $x$ can vary within the range $[(s+1), m]$. Hence, the probability of $e_{m+1}$ being reported as a FN becomes,

$$P_{FNR} = \sum_{x=s+1}^{m} P_{lr} \leq \sum_{x=s+1}^{m} \frac{1}{U.x} \cdot \left[1 - \left(\frac{m-x+1}{m}\right)^k\right]$$

$$= \sum_{x=s+1}^{m} \frac{1}{U.x} - \sum_{x=s+1}^{m} \frac{1}{U.x} \cdot \left(\frac{m-x+1}{m}\right)^k$$

$$= \frac{1}{U} \cdot \ln\left(\frac{m}{s+1}\right) - \frac{1}{U} \cdot \sum_{x=s+1}^{m} \frac{1}{x} \cdot \left(1 - \frac{x-1}{m}\right)^k$$

$$\approx \frac{1}{U} \cdot \ln\left(\frac{m}{s+1}\right) - \frac{1}{U} \cdot \sum_{x=s+1}^{m} \frac{1}{x} \cdot \left(1 - \frac{x.k}{m}\right) \quad [\because \frac{1}{m} \text{ is small}]$$

$$= \frac{1}{U} \cdot \ln\left(\frac{m}{s+1}\right) - \frac{1}{U} \cdot \left[\ln\left(\frac{m}{s+1}\right) - \frac{k.(m-s)}{m}\right]$$

$$\therefore P_{FNR} \leq \frac{k.(m-s)}{U.m} \qquad (5.14)$$

If $e_{m+1}$ had occurred in the first $s$ iterations, then it had definitely been inserted, and $e_{m+1}$ will be a FN if at least one of the bits is 0 at the $(m+1)^{th}$ iteration. The probability that the last insertion of $e_{m+1}$ occurs in the first $s$ iterations is,

$$P_{in\ s} = \left[1 - \left(\frac{U-1}{U}\right)^s\right] \cdot \prod_{i=s+1}^{m} \left[\frac{1}{U} \cdot (1-p_i) + \frac{U-1}{U}\right]$$

$$= \left[1 - \left(\frac{U-1}{U}\right)^s\right] \cdot \prod_{i=s+1}^{m} \left[1 - \frac{p_i}{U}\right]$$

$$\leq \left[1 - \left(\frac{U-1}{U}\right)^s\right] \cdot \prod_{i=s+1}^{m} \left[1 - \frac{s}{mU}\right] \qquad [\because p_i = s/i]$$

$$\approx \left[1 - \left(1 - \frac{s}{U}\right)\right] \cdot \left[1 - \frac{s}{mU}\right]^{m-s} = \frac{s}{U} \cdot e^{-\frac{s}{U.m} \cdot (m-s)}$$

$$\leq \frac{1}{U} \qquad [\text{Using Eq. (5.12)}] \qquad (5.15)$$

Similar to the previous arguments, probability of transition of a bit from 0 to 1 is $1/m$. As the position $y$ can vary within [s+1, m], the probability that the bit is set after $(m+1)^{th}$ iteration is $\left(\frac{m-s}{m}\right)$. Hence for all the Bloom filters, the probability of at least one bit being zero is given by,

$$P_{set} = 1 - \left[1 - \frac{s}{m}\right]^k \approx \frac{s.k}{m} \qquad (5.16)$$

From Eq. (5.15) and (5.16) the FNR in this context is given by $\frac{s.k}{U.m}$. Using Eq. (5.14) and the above result (both can produce an FNR), the probability of $e_{m+1}$ being reported as a FN can be bounded by,

$$P_{FNR} \approx O\left(\frac{k}{U}\right) \qquad (5.17)$$

Hence, $RSBF$ tends to observe a constant $FNR$ as the stream length increases.

## 5.3 Stability Factor

$SBF$ introduced the concept of *stability* of a Bloom filter, whereby the number of 1s or 0s in the structure become constant after a time period. It should be noted that as the FPR and FNR is dependent on the 1s and 0s present in the Bloom filter respectively, stability of their counts nearly guarantees constant performance of the data structure. In the analysis that follows, we show that our $RSBF$ structure attains stability much earlier compared to $SBF$, which guarantees to achieve stability theoretically at infinite stream length.

In the following theorem, we intend to find out the expected fraction of ones in the RSBF. The fraction of ones (or zeroes) is important because the false positive rate (or FNR) is dependent on the fraction of ones (or zeroes). The faster we attain stability, the better will be the overall performance of the structure.

Let $E(X)$ be expected count of 1s in one of the $k$ Bloom filters of RSBF; then the expected fraction of ones in $RSBF$, ($\zeta$) can be approximated by $\frac{E(X)}{s}$, where $s$ is the size of each Bloom filter (in bits).

**THEOREM 5.1.** *Given an RSBF with $k.s$ bits, at any iteration $i$, the expected fraction of ones ($\zeta$) is a constant, $\forall i > s$.*

**PROOF.** Let $\lambda$ denote the count of ones in iteration $(i-1)$. We begin our analysis with a single Bloom filter as other Bloom filters (and the operations on them) are identical. We observe that by Algorithm 1, the count of ones can either increase or decrease by **one** only or remain the same in iteration $i$. Therefore, the expected count of ones can be expressed as,

$$E(X) = (\lambda-1)\Pr(\lambda-1) + \lambda\Pr(\lambda) + (\lambda+1)\Pr(\lambda+1) \quad (5.18)$$

since $\Pr(\lambda \pm j) = 0$, where $j \geq 2$.

The count of ones in a Bloom filter can decrease by one when an element is inserted and the bit selected to be set was already set to 1, and during deletion, one of the set bits is reset to 0. The probability is given by,

$$\Pr(\lambda-1) = p_i \left[\frac{\lambda(\lambda-1)}{s^2}\right] \qquad (5.19)$$

The count of ones can remain the same when the $i^{th}$ element $e_i$ in the stream is not inserted. Further, if the element is inserted, the count of ones can still remain the same if a 0 bit is selected to be set to 1 and a 1 bit is reset to 0 during deletion. Also, if the bit to be set to 1 is already set and that to be reset is already 0, the count of ones remain constant. Hence,

$$\Pr(\lambda) = (1-p_i) + p_i \left[\frac{\lambda(s-\lambda+1)}{s^2} + \frac{\lambda(s-\lambda)}{s^2}\right] \quad (5.20)$$

Similarly, the count can increase by one if a 0 bit is set to 1 and during deletion any 0 bit is selected.

$$\Pr(\lambda+1) = p_i \left(\frac{s-\lambda}{s}\right)^2 \qquad (5.21)$$

Substituting Equations 5.19, 5.20 and 5.21 in Equation 5.18, we have,

$$E(X) = p_i \left[\lambda \left(\frac{1-s}{s}\right)^2 + 1\right] + \lambda(1-p_i)$$

$$= \lambda + p_i \left[ \lambda \left( \left( \frac{s-1}{s} \right)^2 - 1 \right) + 1 \right] = \lambda + p_i . \epsilon \qquad (5.22)$$

For any value that $\lambda$ can assume, we have $0 \leq |\epsilon| \leq 1$ and therefore the fraction of ones, $E(X)/s$ in a buffer is a constant. Moreover, the fraction $|p_i . \epsilon|$ is monotonically decreasing with increasing values of $i$, which is the stream length. Hence for large $i$, $\epsilon$ is practically 0. This analysis holds identically for all the remaining $(k-1)$ buffers. Therefore $\zeta$ is a constant for $RSBF$.  $\square$

We now calculate the variance of the count of ones in a single Bloom filter, $Var[X]$ which can be easily extended to the remaining Bloom filters, as discussed previously. Given,

$$Var[X] = E[X^2] - (E[X])^2$$

by simple algebraic manipulations, we have

$$Var[X] \approx 2 \left( \frac{\lambda}{s} \right) p_i \left[ \frac{\lambda}{s} - 1 \right] + p_i - p_i^2 \qquad (5.23)$$

Let $\lambda = \beta.s$, where $0 \leq \beta \leq 1$. Substituting this value in Eq. (5.23), we have,

$$Var[X] = p_i \left( \beta^2 + (\beta - 1)^2 \right) - p_i^2 \qquad (5.24)$$

Eq. (5.24) implies that the variance of the count of ones in the Bloom filters for $RSBF$ is significantly low. For instance, when $\beta = 0.5$ the variance is only $(p_i/2 - p_i^2)$. Further, as the length of the stream increases, the variance of the number of ones decreases in the Bloom filters for $RSBF$.

This analysis implies a faster convergence to stability for $RSBF$ with respect to $SBF$ which is validated by experimental results provided in Section 6 for different datasets.

## 5.4 Setting of Parameters

We explore the procedure of setting the parameters for the proposed algorithm to optimize its performance. Given a fixed amount of memory space, $M$ in bits, we theoretically search for the best setting of the number of Bloom filters, $k$ and the size of each Bloom filter, $s$, such that $s.k = M$. The algorithm takes $M$ and the threshold FPR, $FPR_t$ as inputs, and computes the optimal value of $k$ and $s$ to find a suitable operating point for $RSBF$ with low overall $FPR$ and $FNR$.

Assume that the algorithm conforms to the threshold FPR, $FPR_t$ after the initial $s$ elements of the stream has been processed. An FPR will occur for an element $e_{s+1}$ if all the corresponding bits in the Bloom filter for $e_{s+1}$ are set. Considering a single Bloom filter, the particular bit into which $e_{s+1}$ hashes to will be set if at least one of the $s$ elements maps into it. Therefore,

$$P_{set} = 1 - P(\text{bit is not set by any of the } s \text{ elements})$$

$$= 1 - \left( 1 - \frac{1}{s} \right)^s \approx \left( 1 - \frac{1}{e} \right) \qquad (5.25)$$

Hence for all bits of $e_{s+1}$ in the Bloom filters, Eq. (5.25) becomes

$$P_{FPR_s} = \left( 1 - \frac{1}{e} \right)^k \qquad (5.26)$$

Also, $e_{s+1}$ should not have occurred in the initial $s$ elements. This can be captured by the factor $((U-1)/U)^s$. Considering $U$ to be large, this factor tends to 1, and hence we ignore this term in the present discussion.

Equating $FPR_t$ and Eq. (5.26) we have

$$\left( 1 - \frac{1}{e} \right)^k = FPR_t$$

$$\therefore k = \frac{\ln (FPR_t)}{\ln \left( 1 - \frac{1}{e} \right)} \qquad (5.27)$$

$$\text{and,} \ s = \frac{M}{k} \qquad (5.28)$$

We find that $FPR$ decreases with increase in the value of $k$, while $FNR$ is the lowest when $k = 1$. Hence, to optimize this trade-offs, we take the value of $k$ as the *arithmetic mean* of 1 and that obtained in Eq. (5.27). Given the value of $k$, $s$ can thus be appropriately set according to Eq. (5.28).

For applications requiring a low $FNR$, we can set $k = 1$, and for low $FPR$ requirements $k$ is set as Eq. (5.27). Hence, the $RSBF$ algorithm can dynamically be suited to a particular application needs. Section 6 exhibits that such choice of parameters help $RSBF$ perform better than the competing algorithms.

## 6. RESULTS & ANALYSIS

We implemented both $RSBF$ and $SBF$ [6] algorithms and compared their performance based on real as well as synthetic datasets. The real dataset containing clickstream data [1] having around 3M records and random dataset with 1B records were used to evaluate the quality of membership query results generated.

We performed two sets of experiments to capture: $(a)$ Variation of FNR, FPR and convergence with increasing number of records in the input, and $(b)$ Variation of FNR and FPR with increasing amounts of memory for sampling the input stream, using multiple datasets for increasing percentage of duplicates. In all the experiments, $p^*$ was set to 0.03. For faster changing streams or for more biased reservoir sampling method, $p^*$ can be set to a higher value.

### 6.1 Quality Comparison

In this section we present the variation of FNR and FPR along with convergence rates with increasing number of records in the input stream. The memory used for the underlying Bloom filter data structure is kept constant for both $SBF$ and $RSBF$ in these experiments. For sake of clarity, points are plotted in the curves at every 1K input stream records.

Fig. 2 presents the comparison of FPR for real dataset with more than 3M records. Initially, till the number of input stream records reaches the threshold, $RSBF$ has better FPR (0.001) than $SBF$ (around 0.0025). $RSBF$ in this stage accepts all the input records in its reservoir and the available memory determines the threshold count. It can also be observed that uptil the threshold point $RSBF$ will not incur any FNR.

As the number of records increase, the FPR performance of $RSBF$ gradually becomes comparable to that of $SBF$. We note here that, even with a small memory of 2KB for around 3M elements, the FPR achieved is quite low, 0.0025. This demonstrates that both $RSBF$ and $SBF$ attain low FPR for large number of records with a significantly small memory space.

Fig. 3 presents the comparison of FPR for the synthetic dataset with 1B records. With 128MB memory, as the number of records increases, the FPR for $RSBF$ stabilizes at 0.8%, while that for $SBF$ stabilizes around 0.7%. With larger memory, 512MB memory, as the number of records increases, the FPR for both $RSBF$ and $SBF$ stabilizes at around 0.06%. Thus both $RSBF$ and $SBF$ attain com-parable FPR for massive number of records, with the performance becoming nearly equal at larger memory. The use of *reservoir sampling* in $RSBF$ enables the data structure in general to sieve out duplicates which occur in higher probability as the stream length increases, given the finite size of alphabet set of the input elements.

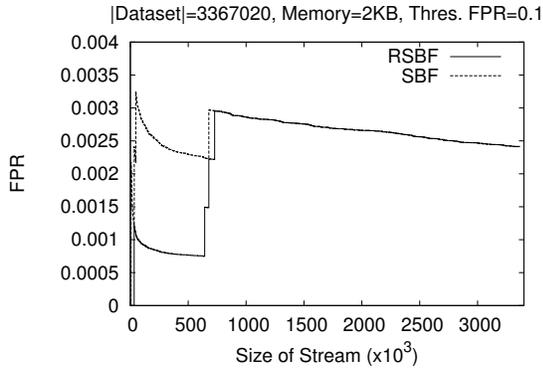

**Figure 2: FPR Comparison**

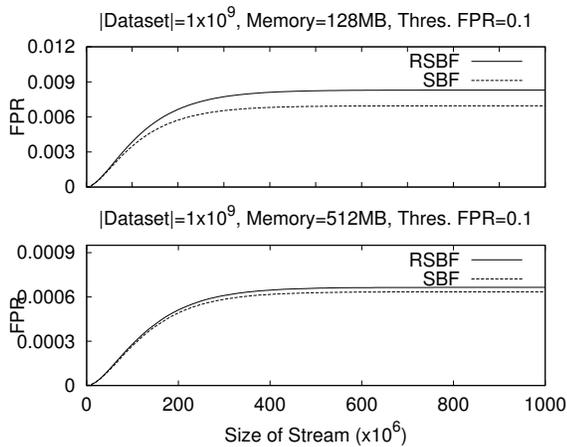

**Figure 3: FPR Comparison**

Fig. 4 compares the FNR between $RSBF$ and $SBF$ with increase in the number of records. For around 3M records and FPR threshold of 0.1, both $RSBF$ and $SBF$ show an initially increases in FNR, stabilizing as the number of records increases further. However, we observe that for both 2KB and 4KB memory, $RSBF$ clearly outperforms $SBF$ by a significant margin. For 2KB memory, $RSBF$ has a stable FNR of 10% which is around $1.5\times$ better as compared to $SBF$ which produces a stable FNR of 15%. With increase in memory the performance gap between the two further increases in favor of $RSBF$. We observe that for 4KB memory, $RSBF$ attains a stable FNR of nearly 12% which is around $1.83\times$ better than that of $SBF$ with a stable FNR of 22%.

Fig. 5 also compares the FNR between $RSBF$ and $SBF$ albeit on synthetic dataset having 1B records and FPR threshold of 0.1. For both $RSBF$ and $SBF$ the FNR again initially increases but then stabilizes. We observe that for both 128MB and 512MB memory, $RSBF$ similarly outperforms $SBF$. For 128MB memory, $RSBF$ has a stable FNR of 22% which is around $1.73\times$ better compared to $SBF$ which has an FNR of 38%. With 512MB memory, for $RSBF$ we observe a stable FNR of 7%, around $1.86\times$ better than $SBF$ with a stable FNR of 13%. Thus, $RSBF$ consistently demonstrates better FNR than $SBF$ upto to a factor of $1.86\times$, for different datasets.

We emphasize that such significant reduction in FNR is novel with respect to stable Bloom filters and extremely vital for practical applications such as search engines. This performance of $RSBF$ can be attributed to the forced insertion of a stream element into the reservoir when the insert probability for the system falls below the threshold $p^*$ as described earlier (Section 4). This approach eliminates the possibility of an FNR occurring due to repeated rejection of an element from being inserted into the reservoir given the lone operation of reservoir sampling. It can also be observed that essentially it helps $RSBF$ to adapt its reservoir in dynamic streaming environments. Hence, it partially acts as a simple bias function for $RSBF$. $SBF$, on the other hand fails to meet such demands.

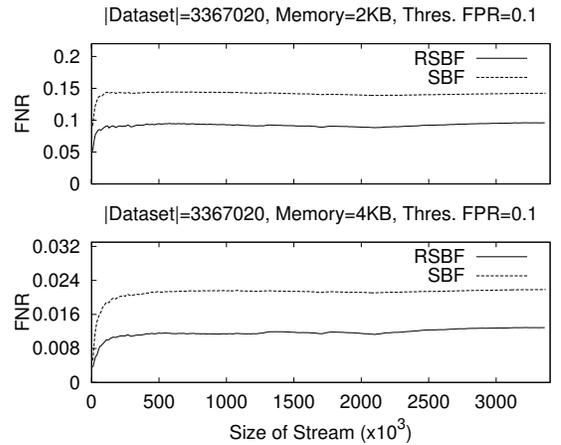

**Figure 4: FNR Comparison: Real Dataset**

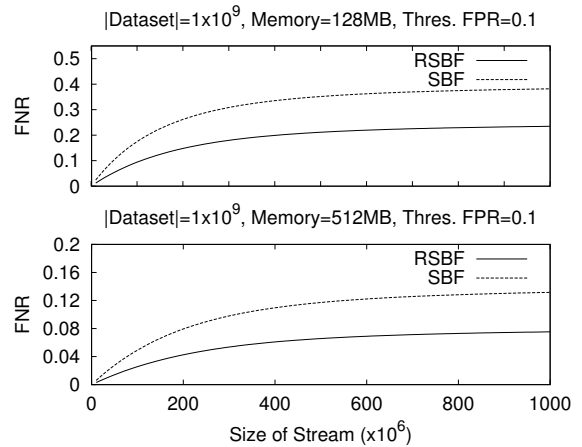

**Figure 5: FNR Comparison: Synthetic Dataset**

[6] proposes $SBF$ having a unique feature, *stability* of the number of 1s present in the Bloom filter leading to a stable performance of $SBF$ in terms of FPR and FNR. This stability poses an attractive feature for applications for guaranteeing a constant performance with increasing stream lengths. However, $SBF$ converges to its stable point at a theoretical stream length of infinity. Practically,

this represents a very large input stream. $RSBF$ also exhibits such stability but converges to a stable performance at a much earlier point. This enables applications to guarantee efficiency at a much smaller stream length.

Fig. 6 compares the difference in the number of $1s$ for successive number of records, in the underlying Bloom filter data structures. By studying the variation in the difference in number of $1s$ with increasing number of records, one gets insights into the convergence behavior of the two algorithm. Here again, the total number of records is around 3M and FPR threshold used is 0.1. For 2KB memory, $RSBF$ stabilizes quickly as the difference in the number of $1s$ stabilizes to nearly 0 at only 500K records. However, $SBF$ does not stabilize even at 3M records. For 4KB memory, $RSBF$ observes stability at around 1.5M records, but $SBF$ fails to stabilize even at 3M records. This demonstrates that our algorithm, $RSBF$ has much better convergence rate than $SBF$.

Fig. 7 similarly compares the difference in the number of $1s$ of successive number of records for the synthetic dataset. With 512KB memory, the difference in the number of $1s$ stabilizes to zero faster for $RSBF$ (shortly after 50 million records) as compared to $SBF$, which has not yet stabilized even at 455 million records. This exactly validates Eq. (5.22) that the number of 1s in $RSBF$ becomes nearly constant much ahead of $SBF$.

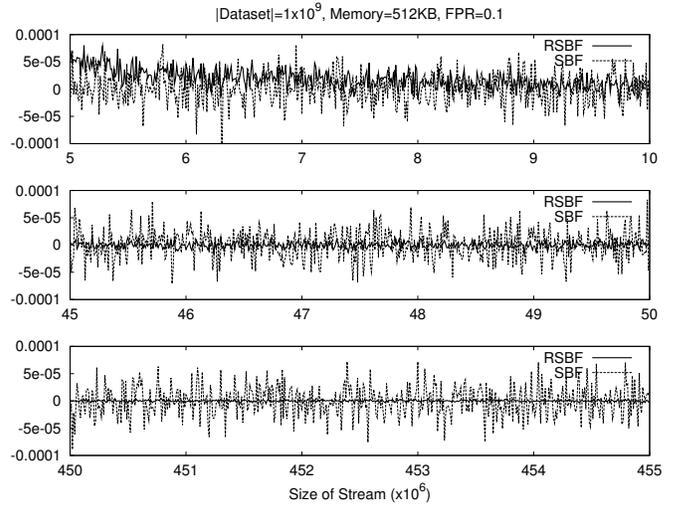

**Figure 7: Convergence Rate Comparison: Synthetic Dataset**

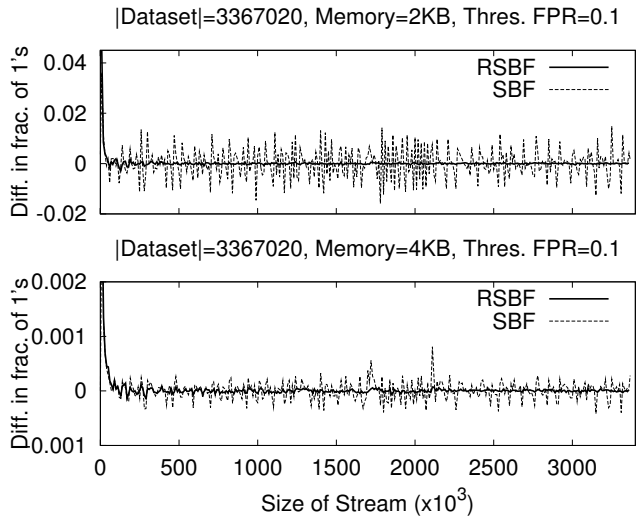

**Figure 6: Convergence Rate Comparison: Real Dataset**

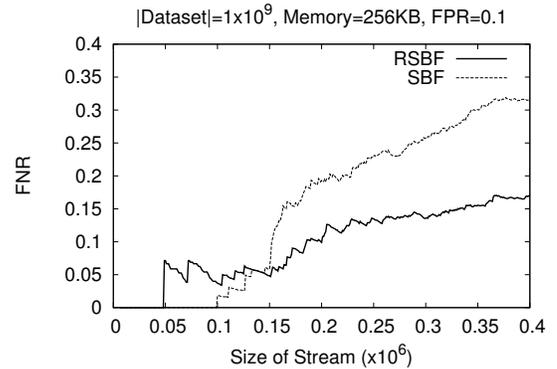

**Figure 8: FNR Stability Rate: Synthetic Dataset**

We further emphasize the faster convergence of $RSBF$ in Fig. 8 which compares the FNR of both the algorithms with increase in stream length. We observe that the increase of FNR in $RSBF$ is around 0.1% over a stream length of 0.35M elements, having an average deviation of $0.3 \times 10^{-6}$ per element. On the other hand, $SBF$ demonstrates an increase in FNR of around 0.3% over 0.3M element with the average deviation as $1 \times 10^{-6}$. Therefore, one can figure out that $RSBF$ converges to an almost stable FNR much earlier in the stream than $SBF$. This in turn reinforces the faster convergence of the stability curves of $RSBF$ compared to $SBF$, described earlier.

## 6.2 Detailed Analysis

In this section, we present detailed analysis of the algorithms, $RSBF$ and $SBF$ compared against variation of memory used and percentage of distinct elements in the stream.

Table 2 presents the FNR and FPR with 100K records and 76%

distinct records while varying memory used for the underlying Bloom Filter data structure from 16K bits to 4.2M bits. Here, both the FNR and FPR of $RSBF$ and $SBF$ are close to each other for different values of memory used. This is due to the fact that the stream size is quite small and neither of the structures have reached their stability point. Table 3 presents the FNR and FPR with 10M records and 49% distinct records with varying memory from 16K bits to 4.2M bits. Here again, we observe comparable results for both FNR and FPR in $RSBF$ and $SBF$ as the number of duplicates and distinct elements in the stream are roughly equal. However, with 10M records, and percentage of distinct elements lesser than 49%, $RSBF$ has better FNR than $SBF$ as exhibited in other dataset values given below.

| Space (in bits) | SBF % FNR | RSBF % FNR | SBF % FPR | RSBF % FPR |
|---|---|---|---|---|
| 16384 | 85.06 | 84.49 | 10.05 | 11.22 |
| 65536 | 74.37 | 74.85 | 8.093 | 8.384 |
| 4194304 | 5.51 | 6.29 | 0.00382 | 0.00263 |

**Table 2: Dataset of 100K elements (76% Distinct)**

| Space (in bits) | SBF % FNR | RSBF % FNR | SBF % FPR | RSBF % FPR |
|---|---|---|---|---|
| 16384 | 88.83 | 87.52 | 11.08 | 12.464 |
| 262144 | 88.11 | 86.89 | 10.86 | 12.12 |
| 4194304 | 77.33 | 77.73 | 7.822 | 7.914 |

**Table 3: Dataset of 10M elements (49% Distinct)**

Table 4 presents the FNR and FPR with 695M records and 15% distinct records while varying memory used for the underlying Bloom Filter data structure from 262K bits to 4.2B bits. Here, FNR achieved by *RSBF* is better than *SBF* and this gap is higher when larger memory is used. At around 67M bits, *RSBF* has FNR of 58.3%, while *SBF* has FNR of 82.48%; while at 1B bits, *RSBF* has FNR of 23.12%, while *SBF* has FNR of 37.79%. However, the FPR values remain similar across both these algorithms.

Table 5 presents the FNR and FPR with 1B records and 10% distinct records while varying memory used for the underlying Bloom Filter data structure from 262K bits to 4.2B bits. Again, FNR achieved by *RSBF* is better than *SBF*. At around 67M bits, *RSBF* has FNR of 58%, while *SBF* has FNR of 82%; while at 1B bits, *RSBF* has FNR of 23.47%, while *SBF* has FNR of 37%. The ratio of FNR between *SBF* and *RSBF* increases to 1.74× at 4.2B bits. However, the FPR values remain similar across both these algorithms. This demonstrates, that our algorithm, *RSBF* has consistent superior FNR compared with *SBF*, with FPR values close to *SBF* though sometimes higher by a small margin.

| Space (in bits) | SBF % FNR | RSBF % FNR | SBF % FPR | RSBF % FPR |
|---|---|---|---|---|
| 262144 | 88.86 | 87.47 | 12.51 | 11.1 |
| 67108864 | 82.48 | 58.2818 | 8.3 | 8.4 |
| 1073741824 | 37.79 | 23.12 | 0.742 | 0.89 |
| 4294967296 | 12.94 | 7.37 | 0.069 | 0.072 |

**Table 4: Dataset of 695M elements (15% Distinct)**

| Space (in bits) | SBF % FNR | RSBF % FNR | SBF % FPR | RSBF % FPR |
|---|---|---|---|---|
| 67108864 | 82.58 | 67.66 | 8.262 | 10.262 |
| 1073741824 | 38.17 | 23.47 | 0.7 | 0.83 |
| 4294967296 | 13.163 | 7.53 | 0.0634 | 0.0664 |

**Table 5: Dataset of 1B elements (10% Distinct)**

## 7. CONCLUSIONS & FUTURE WORK

Real-time data redundancy removal for streaming datasets poses a challenging problem. We have presented the design of a novel Bloom filter based on biased Reservoir Sampling. Using threshold based non-temporal bias function, we obtain improved FNR and convergence rates as compared to [6] while maintaining similar FPR. Using detailed theoretical analysis, we prove upper bounds on FPR and FNR. Further, we prove better convergence (stability of number of 1s) of our algorithm with expected bounds on the number of 1s compared to other algorithms.

We demonstrate real-time in-memory DRR using both real and synthetic datasets of the order of 1B records. We demonstrate upto 2× better FNR and much better convergence rates compared to the best [6] prior results. To the best of our knowledge, *RSBF* offers the best known FNR and convergence rates for streaming datasets with the same memory requirement as that of *SBF*. In future, we hope to study the effect of other temporal and non-temporal biased functions for reservoir sampling to improve the FNR. Investigation for methods of parallelizing the *RSBF* algorithm may in turn lead to further advancements of parallel data redundancy removal research.